\documentclass[12pt]{article}

\usepackage[preprint]{neurips_2023}

\usepackage[utf8]{inputenc} 
\usepackage[T1]{fontenc}    
\usepackage{hyperref}       
\usepackage{url}            
\usepackage{booktabs}       
\usepackage{amsfonts}       
\usepackage{nicefrac}       
\usepackage{microtype}      
\usepackage{xcolor}         

\usepackage{fancyhdr}

\hypersetup{colorlinks, linkcolor=blue, citecolor=blue, urlcolor=blue}

\usepackage{amsfonts,dsfont}
\usepackage{hyperref}
\usepackage{color}
\usepackage{amsmath, amssymb}
\usepackage{amsthm}
\usepackage{bm}
\usepackage{stmaryrd}
\usepackage{color}
\usepackage{comment}
\usepackage{geometry}
\allowdisplaybreaks
\usepackage{tikz}

\usepackage{graphicx}
\usepackage{caption}
\usepackage{subcaption}

\usepackage{tabularx}

\usepackage{caption}

\theoremstyle{definition}\newtheorem{definition}{Definition}
\theoremstyle{definition}\newtheorem{proposition}[definition]{Proposition}
\theoremstyle{definition}\newtheorem{remark}[definition]{Remark}
\theoremstyle{definition}\newtheorem{theorem}[definition]{Theorem}
\theoremstyle{definition}
\theoremstyle{definition}
\theoremstyle{definition}\newtheorem{example}[definition]{Example}
\theoremstyle{definition}
\theoremstyle{definition}
\theoremstyle{definition}

\newcommand{\dt}{\mathrm{d}}
\newcommand{\Dt}{\mathrm{D}}

\newcommand{\mK}{\mathcal{K}}
\newcommand{\n}{^{(n)}}
\newcommand{\pp}[2]{\frac{\partial #1}{\partial #2}}
\newcommand{\vv}{\mathbf{v}}
\newcommand{\tD}{\text{D}}
\newcommand{\K}{\mathcal{K}}
\newcommand{\orho}{\overline{\rho}}

\newcommand{\J}{\mathrm{J}}

\title{Invariant Physics-Informed Neural Networks for Ordinary Differential
Equations}

\author{%
  Shivam Arora \\
  Department of Mathematics and Statistics\\
  Memorial University of Newfoundland\\
  St. John’s, NL, A1C 5S7, Canada \\
  \texttt{sarora17@mun.ca} \\
  \And
  Alex Bihlo \\
  Department of Mathematics and Statistics\\
  Memorial University of Newfoundland\\
  St. John’s, NL, A1C 5S7, Canada \\
  \texttt{abihlo@mun.ca} \\
  \AND
  Francis Valiquette \\
  Department of Mathematics \\
  Monmouth University \\
  West Long Branch, NJ, 07764, USA\\
  \texttt{fvalique@monmouth.edu} \\
}

\begin{document}

\maketitle

\vspace{0.25cm}

\begin{abstract}
Physics-informed neural networks have emerged as a prominent new method for solving differential equations. While conceptually straightforward, they often suffer training difficulties that lead to relatively large discretization errors or the failure to obtain correct solutions.  In this paper we introduce \textit{invariant physics-informed neural networks} for ordinary differential equations that admit a finite-dimensional group of Lie point symmetries.  Using the method of equivariant moving frames, a differential equation is invariantized to obtain a, generally, simpler equation in the space of differential invariants. A solution to the invariantized equation is then mapped back to a solution of the original differential equation by solving the reconstruction equations for the left moving frame.  The invariantized differential equation together with the reconstruction equations are solved using a physics-informed neural network, and form what we call an invariant physics-informed neural network.  We illustrate the method with several examples, all of which considerably outperform standard non-invariant physics-informed neural networks.
\end{abstract}

\section{Introduction}

Physics-informed neural networks (PINNs) are an emerging method for solving differential equations using deep learning, cf.\ \cite{laga98a,rais19a}. The main idea behind this method is to train a neural network as an approximate solution interpolant for the differential equation. This is done by minimizing a loss function that incorporates both the differential equation and its initial and/or boundary conditions. The method has a particular elegance as the derivatives in the differential equation can be computed using automatic differentiation rather than numerical discretization, which greatly simplifies the solution procedure, especially when solving differential equations on arbitrary surfaces, see, e.g.\ \cite{tang22a}. 

The ease of the discretization procedure in physics-informed neural networks, however, comes at the price of numerous training difficulties, and numerical solutions that are either not particularly accurate, or fail to converge at all to the true solutions of the differential equations.  Since training a physics-informed neural network constitutes a non-convex optimization problem, an analysis of failure modes when physics-informed neural networks fail to train accurately is a non-trivial endeavour. This is why several modified training methodologies have been proposed, which include domain decomposition strategies, \cite{jagt20a}, modified loss functions, \cite{wang22a}, and custom optimization, \cite{bihl23a}. While all of these strategies, sometimes substantially, improve upon vanilla physics-informed neural networks, none of these modified approaches completely overcome all the inherent training difficulties. 

Here we propose a new approach for training physics-informed neural networks, which relies on using Lie point symmetries of differential equations and the method of equivariant moving frames to simplify the form of the differential equations that have to be solved.  This is accomplished by first projecting the differential equation onto the space of differential invariants to produce an \emph{invariantized differential equation}.  The solution to the invariantized equation is then mapped back to the solution of the original equation by solving a system of first order differential equations for the left moving frame, called \emph{reconstruction equations}.  The invariant physics-informed neural network architecture proposed in this paper consists of simultaneously solving the invariantized differential equation and the reconstruction equations using a physics-informed neural network.   The method proposed is entirely algorithmic, and can be implemented for any system of differential equations that is strongly invariant under the action of a group of Lie point symmetries. Since many equations of physical relevance admit a non-trivial group of Lie point symmetries, the proposed method is  potentially a viable path for improving physics-informed neural networks for many real-world applications. The idea of projecting a differential equation into the space of invariants and then reconstructing its solution is reminiscent of the recent work by \cite{vaqu23Ay}, where the authors consider Hamiltonian systems with symmetries, although the tools used in our paper and in \cite{vaqu23Ay} to achieve the desired goals are very different. Moreover, in our approach we do not assume that our equations have an underlying symplectic structure.

To simplify the theoretical exposition, we focus on the case of ordinary differential equations in this paper.  We show using several examples that the proposed approach substantially improves upon the numerical results achievable with vanilla physics-informed neural networks.  Applications to partial differential equations will be considered elsewhere.

The paper is organized as follows. We first review relevant work on physics-informed neural networks and symmetry-preserving numerical methods in Section~\ref{sec:PreviousWork}. In Section~\ref{sec:GroupFoliation} we introduce the method of equivariant moving frames and review how it can be used to solve ordinary differential equations that admit a group of Lie point symmetries. Building on Section \ref{sec:GroupFoliation}, we introduce a version of invariant physics-informed neural network in Section~\ref{sec:InvariantPINNs}. We illustrate our method with several examples in Section~\ref{sec:Examples}. The examples show that our proposed invariant physics-informed neural network formulation can yield better numerical results than its non-invariant version.  A short summary and discussion about potential future research avenues concludes the paper in Section~\ref{sec:Conclusion}. 

\section{Previous work}\label{sec:PreviousWork}

Physics-informed neural networks were first proposed in~\cite{laga98a}, and later popularized through the work of \cite{rais19a}. The main idea behind physics-informed neural networks is to train a deep neural network to directly approximate the solution to a system of differential equations. This is done by defining a loss function that incorporates the given system of differential equations, along with any relevant initial and/or boundary conditions. Crucially, this turns training physics-informed neural networks into a multi-task, non-convex optimization problem that can be challenging to minimize, cf.\ \cite{kris21a}. There have been several solutions proposed to overcome the training difficulties and improve the generalization capabilities of physics-informed neural networks.  These include modified loss functions, \cite{mccl20a,wang22a}, meta-learned optimization, \cite{bihl23a}, domain decomposition methods, \cite{bihl22a,jagt20a}, and the use of operator-based methods, \cite{brec23a,lu21a}.

The concepts of symmetries and transformation groups have also received considerable attention in the machine learning community. Notably, the equivariance of convolutional operations with respect to spatial translations has been identified as a crucial ingredient for the success of convolutional neural networks, \cite{cohe16a}. The generalization of this observation for other types of neural network layers and other transformation groups has become a prolific subfield of deep learning since.  For example, see \cite{finz21a} for some recent results.

Here we do not consider the problem of endowing a neural network with equivariance properties but rather investigate the question of whether a better formulation of a given differential equation can help physics-informed neural networks better learn a solution. As we will be using symmetries of differential equations for this re-formulation, our approach falls within the framework of geometric numerical integration, see e.g.\ \cite{blan16a}. The problem of symmetry-preserving numerical schemes, in other words the problem of designing discretization methods for differential equations that preserve the symmetries of differential equations, has been studied extensively over the past several decades, see~\cite{doro91Ay,shok83a,yane73a}, for some early work on the topic. Invariant discretization schemes have since been proposed for finite difference, finite volume, finite element and meshless methods, see, for example, \cite{bihl13Ay,bihl22b,bihl14b,bihl17b,bihl19b,budd01Ay,kim06Ay,olve01Ay,rebe11Ay,rebe15a}.

\section{Method}\label{sec:GroupFoliation}

In this section we introduce the theoretical foundations on which the invariant physics-informed neural network framework is based.  In order to fix some notation, we begin by recalling certain well-known results pertaining to symmetries of differential equations, and refer the reader to \cite{blum10Ay,blum89Ay,hydo00Ay,olve86Ay} for a more thorough exposition. Within the field of group analysis, the use of moving frames to solve differential equations is not well-known.  Therefore, the main purpose of this section is to introduce this solution procedure.  We note that, in contrast to the approach proposed in \cite[Chapter 6]{mans10a}, we avoid the introduction of computational variables. Instead, all computations are based on the differential invariants of the prolonged group action.  Our approach is a simplified version of the algorithm presented in \cite{thom15Ay}, which deals with partial differential equations admitting infinite-dimensional symmetry Lie pseudo-groups. 

\subsection{Invariant differential equations}

As mentioned in the introduction, in this paper we limit our attention to the case of ordinary differential equations.  Thus, given a $(q+1)$-dimensional manifold $M$, with $q\geq 1$, let $\J\n = \J\n(M,1)$ denote the $n$\textsuperscript{th} order (extended) jet bundle consisting of equivalence classes of $1$-dimensional curves $C\subset M$ under the equivalence relation of $n$\textsuperscript{th} order contact.  For a detailed exposition of jet spaces, we refer the reader to \cite[Chapter 4]{olve95Ay}.  Introducing the local coordinates $z = (t,u) = (t,u^1,\ldots,u^q)$ on $M$, we consider $t$ to be the independent variable and $u=(u^1,\ldots,u^q)$ to be the dependent variables.   Accordingly, the $n$\textsuperscript{th} order jet space $\J\n$ is parametrized by $z\n = (t,u\n)$, where $u\n$ denotes all the derivatives $u^\alpha_j = u^\alpha_{t^j}$ of order $0 \leq j \leq n$, with $\alpha = 1,\ldots,q$.

Now, let $G$ be an $r$-dimensional Lie group (locally) acting on $M$:
\begin{equation}\label{G}
(T,U) = Z = g\cdot z = g\cdot(t,u), \qquad \text{where}\qquad g\in G.
\end{equation}
Since group transformations preserve contact, see \cite[Chapter 4]{olve95Ay}, the group action~\eqref{G} induces a prolonged action 
\begin{equation}\label{prolonged action}
Z\n = g\cdot z\n
\end{equation}
on the $n$\textsuperscript{th} order jet space $\J\n$.  Coordinate expressions for the prolonged action \eqref{prolonged action} are obtained by applying the implicit total derivative operator
\[
\Dt_T = \frac{1}{\tD_t(T)} \, \Dt_t,
\qquad\text{where}\qquad
\Dt_t = \pp{}{t} + \sum_{j = 0}^\infty \sum_{\alpha=1}^q\> u^\alpha_{j+1}\pp{}{u^\alpha_j}
\]
denotes the standard total derivative operator, to the transformed dependent variables $U^\alpha$: 
\begin{equation}\label{prolonged action coordinates}
U^\alpha_j = U^\alpha_{T^j} = \Dt_T^j(U^\alpha),\qquad \alpha = 1,\ldots,q,\qquad j \geq 0.
\end{equation} 

We are primarily interested in the action of a Lie group on ordinary differential equations. In the following we use the notation $\Delta(z\n) = \Delta(t,u\n)=0$ to denote a system of differential equations, and use the index notation $\Delta_i(z\n)=0$, $i=1,\ldots,l,$ to label each equation in the system of equations $\Delta(z\n)=0$.  If $\Delta(z\n)=0$ is a single equation, then we omit the indexing notation.

\begin{definition}\label{def: strong invariance}
A nondegenerate\footnote{A differential equation is nondegenerate if at every point in its solution space it is both locally solvable and of maximal rank, \cite[Definition 2.70]{olve86Ay}.} ordinary differential equation $\Delta(z\n)=0$ is said to be \emph{strongly invariant} under the prolonged action \eqref{prolonged action} of a connected local Lie group of transformations $G$ if and only if
\[
\Delta(g\cdot z\n) = 0\qquad \text{for all}\qquad g\in G
\]
near the identity element.
\end{definition}

\begin{remark}
Strong invariance is more restrictive than the usual notion of symmetry, where invariance is only required to hold on the solution space.  In the following, we require strong invariance to guarantee that our differential equation is an invariant function.
\end{remark}

Invariance is usually stated in terms of the infinitesimal generators of the group action.  To this end, let
\begin{equation}\label{v}
\vv_\kappa = \xi_\kappa(t,u)\pp{}{t} + \sum_{\alpha=1}^q \phi^\alpha_\kappa(t,u) \pp{}{u^\alpha},\qquad \kappa = 1,\ldots,r,
\end{equation}
be a basis for the Lie algebra $\mathfrak{g}$ of infinitesimal generators of the group action $G$. The prolongation of the vector fields \eqref{v}, induced from the prolonged action \eqref{prolonged action coordinates}, is given by 
\begin{equation}\label{prv}
\vv\n_\kappa = \xi_\kappa(t,u)\pp{}{t} + \sum_{j=0}^n \sum_{\alpha = 1}^q \phi^{\alpha,j}_\kappa(t,u^{(j)})\pp{}{u^\alpha_j},\qquad \kappa=1,\ldots,r,
\end{equation}
where the coefficients of the prolonged vector fields are computed using the prolongation formula 
\[
\phi^{\alpha,j}_\kappa = \Dt^j_t(\phi^\alpha_\kappa - \xi_\kappa u^\alpha_1) + \xi_\kappa u^\alpha_{j+1},\qquad \kappa=1,\ldots,r,\qquad \alpha=1,\ldots,q,\qquad 0\leq j \leq n,
\]
which can be found in \cite[Eq.\ 2.50]{olve86Ay}. The vector fields \eqref{prv} provide a basis for the Lie algebra of prolonged infinitesimal generators $\mathfrak{g}\n$.

Let $F\colon \J\n \to \mathbb{R}$ be a differential function.  As explained in \cite[Section 1.3]{olve86Ay}, the infinitesimal change of $F$ under the flows generated by the prolonged vector fields \eqref{prv} is given by
\[
\vv_\kappa\n [F(z\n)] = \xi_\kappa \pp{F}{t} + \sum_{j=0}^n \sum_{\alpha = 1}^q \phi^{\alpha,j}_\kappa \pp{F}{u^\alpha_j},\qquad \kappa=1,\ldots,l.
\]
Then, at the infinitesimal level, the strong invariance notion introduced in Definition \ref{def: strong invariance} is equivalent to the following proposition.

\begin{proposition}\label{prop: strong invariance}
A nondegenerate ordinary differential equation $\Delta(z\n)=0$ is strongly invariant under the prolonged action of a connected local Lie group of transformations $G$ if and only if
\[
\vv\n_\kappa[\Delta_i(z\n)]=0,\qquad \kappa=1,\ldots,r,\qquad i=1,\ldots,l,
\]
where $\vv_1,\ldots,\vv_r$ is a basis of infinitesimal generators for the group of transformations $G$.
\end{proposition} 

Proposition \ref{prop: strong invariance} follows from the fact that for a strongly invariant system of differential equations $\Delta(z\n)=0$, each function $\Delta_i(z\n)$, $i=1,\ldots,l$, is a differential invariant function~\cite[Thm 2.8]{olve86Ay}.

\begin{remark}
As one may observe, we do not include the initial conditions
\begin{equation}\label{initial conditions}
u^{(n-1)}(t_0) = u^{(n-1)}_0
\end{equation}
when discussing the symmetry of the differential equation $\Delta(z\n)=\Delta(t,u\n)=0$. This is customary when studying symmetries of differential equations.  Of course, the
 initial conditions are necessary to select a particular solution and when implementing numerical simulations.     
\end{remark} 

We finish the section with two definitions introducing regularity assumptions on the prolonged group action that will guarantee that the solution procedure discussed in the next sections is valid.

\begin{definition}
The (prolonged) group action of $G$ on $\J\n$ is said to be \emph{semi-regular} if all the orbits have the same dimension.  A semi-regular group action is \emph{regular} if, in addition, each point $z\n \in \J\n$ has arbitrarily small neighborhoods whose intersection with each orbit is a connected subset thereof.
\end{definition}

\begin{definition}
The prolonged action of $G$ is said to be \emph{transversed} to the solution space
\[
S\n = \{z\n \in \J\n\,|\, \Delta(z\n)=0\}
\]
of the differential equation $\Delta(z\n)=0$ if at every point $z\n \in S\n$, the intersection of the Lie algebra of prolonged infinitesimal generators $\mathfrak{g}\n|_{z\n}$ at $z\n$ intersects the tangent space $T_{z\n}S\n$ trivially so that $\mathfrak{g}\n \cap TS\n = \{0\}$.
\end{definition}

\subsection{Invariantization}

In this section we assume that $\Delta(z\n)=0$ is a nondegenerate differential equation, which is strongly invariant under the prolonged action of an $r$-dimensional Lie group $G$ acting regularly on $\J\n$.  We now explain how to use the method of equivariant moving frames to ``project" the differential equation onto the space of differential invariants.  For the theoretical foundations of the method of equivariant moving frames, we refer the reader to the foundational papers by \cite{fels99Ay,koga03Ay}, and the textbook by \cite{mans10a}.   

\begin{definition}\label{def: mf}
A \emph{right moving frame} is a map $\rho \colon \J\n \to G$ that satisfies the $G$-equivariance condition
\begin{equation}\label{eq: right equivariance}
\rho(g\cdot z\n) =  \rho(z\n) \cdot g^{-1}
\end{equation}
for all $g\in G$ where the prolonged action is defined.  Taking the group inverse of a right moving frame yields the left moving frame
\[
\orho(z\n) = \rho(z\n)^{-1}
\]
satisfying the equivariance condition
\[
\orho(g\cdot z\n) = g\cdot \orho(z\n).
\]
\end{definition}

To guarantee the existence of a moving frame, we need to introduce the notion of freeness of the (prolonged) group action.

\begin{definition}\label{freeness}
A Lie group $G$ acting on $\J\n$ is said to act \emph{freely} if for all $z\n \in \J\n$ the isotropy group at $z\n$ given by
\[
G_{z\n} = \{g\in G\,|\, g\cdot z\n = z\n \}
\]
is trivial, that is $G_{z\n} = \{e\}$. The Lie group $G$ is said to act \emph{locally freely} if $G_{z\n}$ is a discrete subgroup of $G$ for all $z\n\in \J\n$. 
\end{definition}

\begin{theorem}\label{thm:mf existence}
A moving frame exists in some neighborhood of a point $z\n\in \J\n$ if and only if the prolonged action of $G$ is (locally) free and regular near $z\n$.
\end{theorem}

The proof of Theorem \ref{thm:mf existence} can be found in \cite[Thm 4.4]{fels99Ay}.  

\begin{remark}
In general, Theorem \ref{thm:mf existence} might only hold on a $G$-invariant submanifold $\mathcal{V}\n \subset \J\n$.  In this case we would restrict Definitions \ref{def: mf} and \ref{freeness} to $\mathcal{V}\n$.  To simplify the discussion, we assume that $\mathcal{V}\n = \J\n$ in the subsequent considerations.
\end{remark}

A moving frame is obtained by selecting a (regular) cross-section $\mK\subset \J\n$ to the orbits of the prolonged action.  

\begin{definition}
A (local) \emph{cross-section} is a submanifold $\mathcal{K} \subset \J\n$ of codimension $r=\text{dim }G$ such that $\mathcal{K}$ intersects each orbit transversally.  The cross-section is \emph{regular} if $\mathcal{K}$ intersects each orbit at most once.
\end{definition}

Keeping with most applications, and to simplify the exposition, we assume $\mK$ is a regular coordinate cross-section obtained by setting $r$ coordinates of the jet $z\n$ to constant values:
\begin{equation}\label{coordinate cross-section}
z^{a_\kappa} = c^\kappa,\qquad \kappa=1,\ldots,r.
\end{equation}
Under the assumptions of Theorem \ref{thm:mf existence}, the right moving frame at $z\n$ is the unique group element mapping $z\n$ onto the cross-section $\mathcal{K}$ specified by \eqref{coordinate cross-section}.  This transformation is obtained by solving the \emph{normalization equations}
\[
Z^{a_\kappa} = g\cdot z^{a_\kappa} = c^\kappa,\qquad \kappa=1,\ldots,r,
\]
for the group parameters $g=(g^1,\ldots,g^r)$, yielding the right moving frame $\rho$.  Given a right moving frame, there is a systemic procedure for constructing differential invariant functions.

\begin{definition}\label{def: invariantization}
Let $\rho\colon \J\n \to G$ be a right moving frame.  The invariantization of the differential function $F\colon \J\n \to \mathbb{R}$ is the differential invariant function
\begin{equation}\label{eq: invariantization}
\iota(F)(z\n) = F(\rho(z\n)\cdot z\n).
\end{equation}
\end{definition}

\begin{remark}
The fact that \eqref{eq: invariantization} is a differential invariant function follows from the $G$-equivariant property \eqref{eq: right equivariance} for the right moving frame.
\end{remark}

Applying the invariantization map $\iota$ introduced in Definition \ref{def: invariantization} to $z\n$ componentwise yields the differential invariants
\[
\iota(z\n) = \rho(z\n)\cdot z\n,
\]
which can be used as coordinates for the cross-section $\mathcal{K}$.  In particular, the invariantization of the coordinates used to define the cross-section in \eqref{coordinate cross-section} are constant
\[
\iota(z^{a_\kappa}) = c^\kappa,\qquad \kappa=1,\ldots,r,
\]
and are called \emph{phantom invariants}.  The remaining invariantized coordinates are called \emph{normalized invariants}. 

In light of Theorem 5.32 in \cite{olve09Ay}, assume there are $q+1$ independent normalized invariants
\begin{equation}\label{I}
H, I^1,\,\ldots,I^q,
\end{equation}
such that locally 
\[
I^\alpha = I^\alpha(H),\qquad \alpha = 1,\ldots,q,
\]
are functions of the invariant $H$, and generate the algebra of differential invariants.  This means that any differential invariant function can be expressed in terms of \eqref{I} and their invariant derivatives with respect to $\Dt_H$.  In the following we let $I\n$ denote the derivatives of $I=(I^1,\ldots,I^q)$ with respect to $H$, up to order $n$.

Assuming the differential equation $\Delta(z\n)=0$ is strongly invariant and its solutions are transverse to the prolonged action, this equation, once invariantized, yields a differential equation in the space of differential nvariants
\begin{subequations}\label{inv initial value problem}
\begin{equation}\label{inv eq}
\iota(\Delta)(t,u\n) = \Delta_{\text{Inv}}(H,I^{(k)}) = 0,\qquad \text{where}\qquad k\leq n.
\end{equation}
Initial conditions for \eqref{inv eq} are obtained by invariantizing \eqref{initial conditions} to obtain
\begin{equation}\label{inv initial conditions}
I^{(k-1)}(H_0) = I^{(k-1)}_0.
\end{equation}
\end{subequations}

\begin{example}\label{Schwarz ex1}
To illustrate the concepts introduced thus far, we use Schwarz' equation
\begin{equation}\label{Schwarz}
\frac{u_{ttt}}{u_t} - \frac{3}{2}\bigg(\frac{u_{tt}}{u_t}\bigg)^2 = F(t),
\end{equation}
where, for simplicity, we assume that $F(t)$ is a continuous function. The general solution to \eqref{Schwarz} can be found in \cite[Section 10]{hille76Ay}.   According to \cite{ovsi09Ay}, the Schwarz derivative $\{u,t\} = u_{ttt}/u_t - (3/2)(u_{tt}/u_t)^2$ first appeared in the treatise by \cite{lagr79Ay}. Over time the Schwarz derivative has found applications in complex analysis, one-dimensional dynamics, Teichm\"uller theory, integrable systems, and conformal field theory. Equation \eqref{Schwarz} admits a three-dimensional Lie group of point transformations given by
\begin{equation}\label{SL2 action}
T = t,\qquad U = \frac{\alpha u + \beta}{\gamma u + \delta},\qquad\text{where}\qquad g=\begin{bmatrix}
\alpha & \beta \\ \gamma & \delta
\end{bmatrix} \in \text{SL}(2,\mathbb{R}),
\end{equation}
so that $\alpha\delta-\beta \gamma = 1$.  A cross-section to the prolonged group action 
\begin{align*}
U_T &= \Dt_t(U) = \frac{u_t}{(\gamma u + \delta)^2}, \\
U_{TT} &= \Dt_t(U_T) = \frac{u_{tt}}{(\gamma u+\delta)^2} - \frac{2\gamma u_t^2}{(\gamma u+\delta)^3}, \\
U_{TTT} &= \Dt_t(U_{TT}) = \frac{u_{ttt}}{(\gamma u+ \delta)^2} - \frac{6\gamma u_t u_{tt}}{(\gamma u+\delta)^3} + \frac{6\gamma^2u_t^3}{(\gamma u+\delta)^4},
\end{align*}
is given by
\begin{equation}\label{Schwarz K}
\K = \{u=0,\, u_t = \sigma,\, u_{tt} = 0\}\subset \mathcal{V}\n \subset \J\n,
\end{equation}
where $\sigma = \text{sign}(u_t)$ and $\mathcal{V}\n = \{ z\n \in \J\n\,|\, u_t\neq 0\}$ with $n\geq 2$. Solving the normalization equations
\[
U = 0,\qquad U_{T} = \sigma,\qquad U_{TT} = 0,
\]
for the group parameters, taking into account the unitary constraint $\alpha \delta - \beta \gamma = 1$, yields the right moving frame
\begin{equation}\label{mf}
\alpha = \pm \frac{1}{\sqrt{|u_t|}},\qquad
\beta = \mp \frac{u}{\sqrt{|u_t|}},\qquad 
\gamma = \pm \frac{u_{tt}}{2|u_t|^{3/2}},\qquad
\delta = \pm \frac{2u_t^2-uu_{tt}}{2|u_t|^{3/2}},
\end{equation}
where the sign ambiguity comes from solving the normalization $U_T = \sigma$, which involves the quadratic term $(\gamma u + \delta)^2$.  Invariantizing the third order derivative $u_{ttt}$ produces the normalized differential invariant
\begin{equation}\label{Schwarz derivative}
\iota(u_{ttt}) =  \frac{u_{ttt}}{(\gamma u+ \delta)^2} - \frac{6\gamma u_t u_{tt}}{(\gamma u+\delta)^3} + \frac{6\gamma^2u_t^3}{(\gamma u+\delta)^4}\bigg|_{\eqref{mf}} 
= \sigma\bigg(\frac{u_{ttt}}{u_t} - \frac{3}{2}\bigg(\frac{u_{tt}}{u_t}\bigg)^2\bigg).
\end{equation}
In terms of the general theory previously introduced, we have the invariants
\begin{equation}\label{Schwarz HI}
H=t,\qquad I = \frac{u_{ttt}}{u_t} - \frac{3}{2}\bigg(\frac{u_{tt}}{u_t}\bigg)^2.
\end{equation}
Since the independent variable $t$ is an invariant, instead of using $H$, we use $t$ in the following computations.  The invariantization of Schwarz' equation \eqref{Schwarz} yields the algebraic equation
\begin{equation}\label{I sol}
I = F(t).
\end{equation}
Since the prolonged action is transitive on the fibers of each component $\{(t,u,u_t,u_{tt})\,|\, u_t>0\} \cup \{(t,u,u_t,u_{tt})\,|\, u_t<0\} = \mathcal{V}^{(2)}$, any initial conditions
\begin{equation}\label{Schwarz initial conditions}
u(t_0) = u_0,\qquad u_t(t_0) = u_t^0,\qquad u_{tt}(t_0) = u_{tt}^0,
\end{equation}
is mapped, via the invariantization map $\iota$, to the identities
\[
0 = 0,\qquad \sigma = \sigma,\qquad 0=0.
\]
\end{example}

\subsection{Recurrence relations}\label{sec: recurrence relations}

In this section we introduce the recurrence relations for the normalized differential invariants, and explain how the invariantized equation \eqref{inv eq} can be derived symbolically, without requiring the coordinate expressions for the right moving frame $\rho$ or the invariants $(H,I\n)$.  

A key observation is that the invariantization map $\iota$ and the exterior differential, in general, do not commute
\[
\iota \circ \dt \neq \dt \circ \iota.
\]
The extent to which these two operations fail to commute is encapsulated in the recurrence relations.  To state these equations we need to introduce the (contact) invariant one-form
\[
\varpi = \iota(\dt t) = \rho^*(\Dt_t(T))\, \dt t,
\]
which comes from invariantizing the horizontal one-form $\dt t$, and where $\rho^*$ denotes the right moving frame pull-back.  We refer the reader to \cite{koga03Ay} for more details.

Given a Lie group $G$, assume its elements $g\in G$ are given by a faithful representation.  Then the \emph{right Maurer--Cartan form} is given by
\begin{equation}\label{mc form}
\mu = \dt g\cdot g^{-1}.
\end{equation}
The pull-back of the Maurer--Cartan form \eqref{mc form} by a right moving frame $\rho$ yields the invariant matrix
\begin{equation}\label{inv mc form}
\nu = \dt \rho \cdot \rho^{-1} = \big[I_{ij}\big]\,\varpi,
\end{equation}
where the invariants $I_{ij}$ are called Maurer--Cartan invariants.

\begin{proposition}
Let $F\colon \J\n \to \mathbb{R}$ be a differential function and $\rho\colon \J\n \to G$ a right moving frame.  We then have the recurrence relation
\begin{equation}\label{invariantized recurrence relation}
\dt[\iota(F)] = \iota[\dt F] + \sum_{\kappa = 1}^r \iota[\vv_\kappa\n(F)]\, \nu^\kappa,
\end{equation}
where $\nu^1, \ldots, \nu^r$ is a basis of normalized Maurer--Cartan forms extracted from \eqref{inv mc form}.
\end{proposition}

In particular, substituting for $F$ in \eqref{invariantized recurrence relation} the jet coordinates $z\n = (t,u\n)$ yields the recurrence relations
\begin{equation}\label{recurrence relations}
\begin{aligned}
\dt[\iota(t)] &= \varpi + \sum_{\kappa = 1}^r \iota(\xi_\kappa)\nu^\kappa,\\ 
\dt[\iota(u^\alpha_j)] &= \iota(u^\alpha_{j+1})\varpi + \sum_{\kappa = 1}^r \iota(\phi^{\alpha,j}_\kappa) \nu^\kappa,
\end{aligned}
\end{equation}
where $\xi_\kappa$, $\phi^{\alpha,j}_\kappa$ are the coefficients of the prolonged vector fields in \eqref{prv}.  The recurrence relations for the jet coordinates \eqref{coordinate cross-section} specifying the  coordinate cross-section $\mathcal{K}$ lead to $r$ linear equations for the normalized Maurer--Cartan forms $\nu^1,\ldots, \nu^r$.  Solving those equations and substituting the result back in \eqref{recurrence relations} yields symbolic expressions for the differential of the normalized invariants without requiring the coordinate expressions for the moving frame $\rho$ or $\iota(t)$, $\iota(u\n)$.  More generally, substituting the expressions for the normalized Maurer--Cartan forms  in \eqref{invariantized recurrence relation} gives the symbolic expression for the differential of any invariantized differential function $F$.

\begin{example}
Continuing Example \ref{Schwarz ex1}, we differentiate the group action \eqref{SL2 action} with respect to the group parameters $\alpha$, $\beta$, $\gamma$, respectively, recalling the unitary constraint $\alpha \delta - \beta\gamma=1$, and evaluate the result at the identity transformation $\alpha=\delta=1$, $\gamma = \beta = 0$ to obtain the basis of infinitesimal generators
\begin{equation}\label{sl2 basis}
\vv_1 = \pp{}{u},\qquad \vv_2 = u\pp{}{u},\qquad \vv_3 = u^2 \pp{}{u}.
\end{equation}
The prolongation of those vector fields, up to order 2, is given by
\begin{equation}\label{sl2 prv}
\begin{aligned}
\vv_1^{(3)} &= \pp{}{u},\\
\vv_2^{(3)} &= u \pp{}{u} + u_t \pp{}{u_t} + u_{tt}\pp{}{u_{tt}}, \\
\vv_3^{(3)} &= u^2\pp{}{u} + 2uu_t \pp{}{u_t} + 2(u_t^2 + uu_{tt})\pp{}{u_{tt}}.
\end{aligned}
\end{equation}
Computing the recurrence relations \eqref{recurrence relations} for $t$, $u$, $u_t$, and $u_{tt}$ yields
\begin{equation}\label{sl2 recurrence relations}
\begin{aligned}
\dt[\iota(t)] &= \varpi,\\
\dt[\iota(u)] &= \iota(u_t) \varpi + \nu^1 + \iota(u) \nu^2 + \iota(u)^2 \nu^3,\\
\dt[\iota(u_t)] &= \iota(u_{tt}) \varpi + \iota(u_t) \nu^2+2 \iota(u) \iota(u_t) \nu^3,\\
\dt[\iota(u_{tt})] &= \iota(u_{ttt}) \varpi + \iota(u_{tt})\nu^2 + 2[\iota(u_t)^2+\iota(u)\iota(u_{tt})]\nu^3.
\end{aligned}
\end{equation}
We note that the coefficients of the correction terms involving the normalized Maurer--Cartan forms $\nu^1, \nu^2, \nu^3$ in the recurrence relations \eqref{sl2 recurrence relations} are obtained by invariantizing the coefficients of the prolonged vector fields \eqref{sl2 prv}.

Recalling the cross-section \eqref{Schwarz K} and the invariants \eqref{Schwarz derivative}, \eqref{Schwarz HI}, we make the substitutions $\iota(t)=t$, $\iota(u)=0$, $\iota(u_t) = \sigma$, $\iota(u_{tt})=0$, $\iota(u_{ttt}) = \sigma I$ into \eqref{sl2 recurrence relations} and obtain
\[
\dt t = \varpi,\qquad 0 = \varpi + \nu^1,\qquad
0 = \nu^2,\qquad 0 = I\sigma\, \varpi+2\nu^3.
\]
Solving the last three equations for the normalized Maurer--Cartan forms yields
\[
\nu^1= -\sigma\, \varpi,\qquad \nu^2 = 0,\qquad \nu^3 = -\frac{I\sigma}{2}\varpi.
\]
In matrix form we have
\begin{equation}\label{sl2 mc}
\nu = \begin{bmatrix}
2\nu^2 & \nu^1 \\ -\nu^3 & -2\nu^2
\end{bmatrix} =
 \begin{bmatrix}
0 & - \sigma \\
\frac{1}{2}\sigma I & 0
\end{bmatrix} \varpi = \begin{bmatrix}
0 & - \sigma \\
\frac{1}{2}\sigma F(t) & 0 
\end{bmatrix}\varpi, 
\end{equation}
where we used the algebraic relationship \eqref{I sol} originating from the invariantization of Schwarz' equation \eqref{Schwarz}. The, perhaps, unexpected coefficients in the Maurer--Cartan matrix \eqref{sl2 mc}, namely $2\nu^2$ and $-\nu^3$, originate from the fact that when introducing the basis of infinitesimal generators \eqref{sl2 basis} we scaled $\vv_2$ by $1/2$ and $\vv_3$ by $-1$.
\end{example}

\subsection{Reconstruction}

Let $I(H)$ be a solution to the invariantized differential equation \eqref{inv eq} with initial conditions \eqref{inv initial conditions}. In this section we explain how to reconstruct the solution to the original equation $\Delta(x,u\n)=0$ with initial conditions \eqref{initial conditions}.  To do so, we introduce the reconstruction equations for the left moving frame $\orho = \rho^{-1}$ given by
\begin{equation}\label{reconstruction eq}
\dt \orho= -\orho\cdot\dt \rho \cdot \orho = - \orho\, \nu,
\end{equation}
where $\nu$ is the normalized Maurer--Cartan form introduced in \eqref{inv mc form}. As we have seen in Section \ref{sec: recurrence relations}, the invariantized Maurer--Cartan matrix $\nu$ can be obtained symbolically using the recurrence relations for the phantom invariants.  On the other hand, the Maurer--Cartan invariants $I_{ij}$ in \eqref{inv mc form} can be expressed in terms of $H$, the solution $I(H)$ to the invariantized initial value problem \eqref{inv initial value problem}, and its derivatives.  Thus, equation \eqref{reconstruction eq} yields a first order system of differential equations for the group parameters depending on the independent variable $H$.  Integrating \eqref{reconstruction eq}, we obtain the left moving frame that sends the invariant curve $(H,I(H))$ to the original solution 
\begin{equation}\label{xu sol}
(t(H),u(H)) = \orho(H) \cdot \iota(t,u)(H).
\end{equation}
The transversality of the prolonged group action implies that the derivative $t_H \neq 0$. Assuming $t_H>0$, the initial conditions to the reconstruction equations \eqref{reconstruction eq} are given by 
\begin{equation}\label{orho initial conditions}
\orho(H_0) = \orho_0\qquad \text{such that}\qquad \orho_0 \cdot \iota(t_0,u^{(n-1)}_0) = (t_0,u^{(n-1)}_0).
\end{equation}
If $t_H<0$, one can always reparametrize the solution so that the derivative becomes positive.

The solution \eqref{xu sol} is a parametric curve with the invariant $H$ serving as parameter.  From a numerical perspective, this is sufficient to graph the solution.  Though, we note that since $t_H\neq 0$, it is possible, by the implicit function theorem, to invert $t=t(H)$ to express the invariant $H=H(t)$ in terms of $t$ and recover the solution $u = u(H(t))$ as a function of~$t$.

\begin{example}
The left moving frame 
\[
\orho = \begin{bmatrix}
\alpha & \beta \\ \gamma & \delta
\end{bmatrix}\quad \in \quad \text{SL}(2,\mathbb{R})
\]
that will send the invariant solution \eqref{I sol} to the original solution $u(t)$ of Schwarz' equation \eqref{Schwarz} with initial conditions \eqref{Schwarz initial conditions} must satisfy the reconstruction equations
\begin{subequations}\label{Schwarz initial value problem}
\begin{equation}\label{Schwarz reconstruction equations}
\begin{bmatrix}
\alpha_t & \beta_t \\
\gamma_t & \delta_t
\end{bmatrix}
 = - \begin{bmatrix}
\alpha & \beta \\
\gamma & \delta
\end{bmatrix} \begin{bmatrix}
0 & -\sigma \\
\frac{1}{2}\sigma F(t) & 0 
\end{bmatrix} = 
\begin{bmatrix}
\alpha & \beta \\
\gamma & \delta
\end{bmatrix} \begin{bmatrix}
0 & \sigma \\
-\frac{1}{2}\sigma F(t) & 0 
\end{bmatrix},
\end{equation}
with the initial conditions 
\begin{equation}\label{Schwarz reconstruction eq initial cond}
\delta_0 = \pm\frac{1}{\sqrt{|u_t^0|}},\qquad
\beta_0 = \pm \frac{u_0}{\sqrt{|u_t^0|}},\qquad
\gamma_0 = \mp \frac{u_{tt}^0}{2(u_t^0)^{3/2}},\qquad
\alpha_0 = \pm \sqrt{|u_t^0|} \mp \frac{u_0 u_{tt}^0}{2(|u_t^0|)^{3/2}}.
\end{equation}
\end{subequations}
Once the reconstruction equations \eqref{Schwarz initial value problem} are solved, the solution to Schwarz' equation \eqref{Schwarz} is
\begin{equation}\label{Schwarz sol}
u(t) = \orho\cdot 0 = \frac{\beta}{\delta}.
\end{equation}
\end{example}

\subsection{Summary}\label{summary}

Let us summarize the algorithm for solving an ordinary differential equation $\Delta(t,u\n)=0$ admitting a group of Lie point symmetries $G$ using the method of moving frames.
\begin{enumerate}
\item Select a cross-section $\mathcal{K}$ to the prolonged action.
\item Choose $q+1$ independent invariants $H$, $I^1,\ldots, I^q$ from $\iota(t,u\n)$ that generate the algebra of differential invariants, and assume $I^1(H), \ldots, I^q(H)$ are functions of $H$.
\item Invariantize the differential equation $\Delta(t,u\n)=0$ and use the recurrence relations \eqref{recurrence relations} to write the result in terms of $H$ and $I^{(k)}$ to obtain the differential equation $\Delta_{\text{Inv}}(H,I^{(k)})=0$.
\item Solve the equation $\Delta_{\text{Inv}}(H,I^{(k)})=0$ subject to the initial conditions \eqref{inv initial conditions}.
\item A parametric solution to the original equation $\Delta(t,u\n)=0$ is given by $\orho(H)\cdot \iota(t,u)(H)$, where the left moving frame $\orho(H)$ is a solution of the reconstruction equation \eqref{reconstruction eq} subject to the initial conditions \eqref{orho initial conditions}.
\end{enumerate}

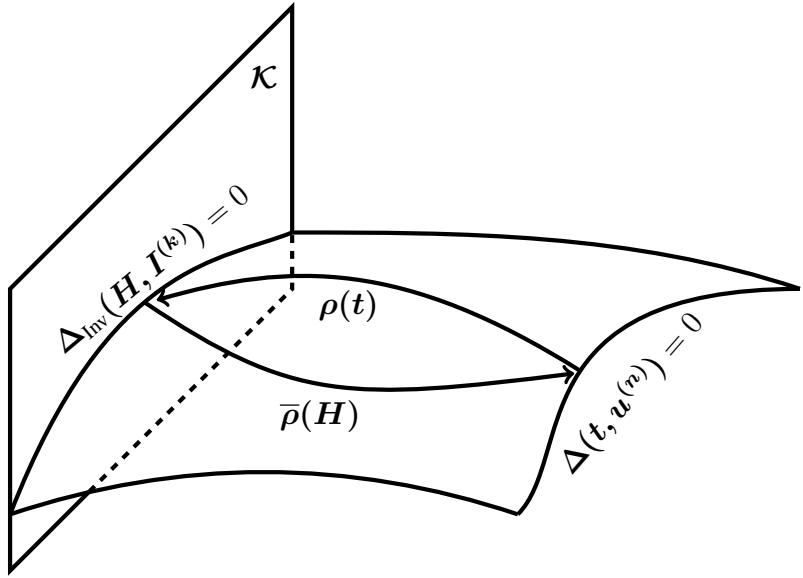
\begin{figure}[!h]
\begin{center}
\begin{tikzpicture}[scale=0.75]
\draw[ultra thick] (1.4,1.4) -- (0,0) -- (0,5) -- (5,10) -- (5,6);
\draw[ultra thick, dashed] (1.4,1.4) -- (5,5) -- (5,6);
\draw[ultra thick] (0,1) .. controls (1.75,5.5) and (3.75,5.5) .. (5,6);
\draw[ultra thick] (0,1) .. controls (3,2) and (6,2) .. (9,1);
\draw[ultra thick] (5,6) .. controls (8,6) and (11,6) .. (14,5);
\draw[ultra thick] (9,1) .. controls (10,2) and (9,5) .. (14,5);
\node at (4.5,8.75) {$\bm{\mathcal{K}}$};
\node[rotate=40] at (2.5,5.35) {$\bm{\Delta_{\text{Inv}}(H,I^{(k)})}=0$};
\node[rotate=49] at (11,3.15) {$\bm{\Delta(t,u^{(n)})}=0$};
\draw[ultra thick,->] (2.4,4.75) .. controls (5,3) and (6,3) .. (10,3.5);
\node at (5.5,2.75) {$\bm{\orho(H)}$};
\draw[ultra thick,->] (10.1,3.55) .. controls (7,5.5) and (5,5.5) .. (2.6,4.8);
\node at (6,4.7) {$\bm{\rho(t)}$};
\end{tikzpicture}
\end{center}
\caption{Solving a differential equation using moving frames.}
\end{figure}

\section{Invariant physics-informed neural networks}\label{sec:InvariantPINNs}

Before introducing our invariant physics-informed neural network, we recall the definition of the standard physics-informed loss function that needs to be minimized when solving ordinary differential equations. To this end, assume we want to solve the ordinary differential equation $\Delta(t, u\n)=0$ subject to the initial conditions $u^{(n-1)}(t_0)=u^{(n-1)}_0$ on the interval $[t_0,t_f]$.  We recall that the ordinary differential equation $\Delta(t,u\n)=0$ can also be a system of differential equations.

First we introduce the collocation points $\{t_i\}_{i=0}^N$ sampled randomly over the interval $[t_0,t_f]$ with $t_0=t_0< t_1 < \cdots < t_N = t_f$.  Then, a neural network of the form $u_{\boldsymbol{\theta}}(t)= \mathcal{N}_{\boldsymbol{\theta}}(t)$, parameterized by the parameter vector $\boldsymbol{\theta}$, is trained to approximate the solution of the differential equation, i.e.\ $u_{\boldsymbol{\theta}}(t)\approx u(t)$, by minimizing the physics-informed loss function
\begin{subequations}\label{eq:PINNloss}
\begin{equation}
\mathcal{L}(\boldsymbol{\theta}) = \mathcal{L}_{\Delta}(\boldsymbol{\theta}) + \alpha\,\mathcal{L}_{\text{I.C.}}(\boldsymbol{\theta})
\end{equation}
with respect to $\boldsymbol{\theta}$, where
\begin{equation}
\mathcal{L}_{\Delta}(\boldsymbol{\theta}) = \sum_{i=0}^N \> \big\|\Delta(t_i, u^{(n)}_{\boldsymbol{\theta}}(t_i))\big\|^2_{\ell^2}
\end{equation}
is the \textit{differential equation loss} and $\|\cdot\|_{\ell^2}$ is the $\ell^2$-norm,
\begin{equation}
\mathcal{L}_{\text{I.C.}}(\boldsymbol{\theta}) = \big\|u_{\boldsymbol{\theta}}^{(n-1)}(t_0)-u^{(n-1)}_0\big\|^2_{\ell^2}
\end{equation}
\end{subequations}
is the \textit{initial condition loss}, and $\alpha \in \mathbb{R}^+ $ is a hyper-parameter to re-scale the importance of both loss functions. We note that the differential equation loss is the mean squared error of the differential equation evaluated at the collocation points $\{t_i\}_{i=0}^N \subset [t_0, t_f]$ over which the numerical solution is sought. The initial condition loss is likewise the mean squared error between the true initial conditions and the initial conditions approximated by the neural network. We note in passage that the initial conditions could alternatively be enforced as a hard constraint in the neural network, see, e.g.\ \cite{brec23a,laga98a}, in which case the physics-informed loss function would reduce to $\mathcal L_{\Delta}(\boldsymbol{\theta})$ only.

The physics-informed loss function \eqref{eq:PINNloss} is minimized using gradient descent, usually using the Adam optimizer, \cite{king14a}, but also more elaborate optimizers can be employed, \cite{bihl23a}. The particular elegance of the method of physics-informed neural networks lies in the fact that the derivatives $u_{\boldsymbol{\theta}}^{(n)}$ of the neural network solution approximation are computed using \textit{automatic differentiation}, \cite{bayd18a}, which is built into all modern deep learning frameworks such as \texttt{JAX}, \texttt{PyTorch}, or \texttt{TensorFlow}.

Similar to the above standard physics-informed neural network, an invariant physics-informed neural network is a feed-forward neural network approximating the solution of the invariantized differential equation and the reconstruction equations for the left moving frame. 

In light of the five step process given in Section \ref{summary}, assume the invariantized equation $\Delta_{\text{Inv}}(H,I^{(k)})=0$ and the reconstruction equation $\dt \orho = -\orho \nu$ have been derived.  Introduce an interval of integration $[H_0,H_f]$ over which the numerical solution is sought, and consider the collocation points $\{H_i\}_{i=0}^N \subset [H_0,H_f]$, such that $H_N=H_f$.   The neural network has to learn a mapping between $H$ and the functions
\[
I_{\boldsymbol{\theta}}(H)\qquad\text{and}\qquad \orho_{\boldsymbol{\theta}}(H),
\]
where $I_{\boldsymbol{\theta}}(H)$ denotes the neural network approximation of the differential invariants $I(H)$ solving \eqref{inv eq}, and $\orho_{\boldsymbol{\theta}}(H)$ is the approximation of the left moving frame $\orho(H)$ solving the reconstruction equations \eqref{reconstruction eq}.  We note that the output size of the network depends on the numbers of invariants $I(H)$ and the size of the symmetry group via $\orho(H)$.

The network is trained by minimizing the invariant physics-informed loss function consisting of the invariantized differential equation loss and the reconstruction equations loss defined as the sum of mean squared errors
\begin{equation}\label{lost function}
\mathcal{L}_{\Delta_{\text{Inv}},\orho}(\boldsymbol{\theta}) = \sum_{i=0}^N \big(\big\|\Delta_{\text{Inv}}(H_i,I^{(k)}_{\boldsymbol{\theta}}(H_i))\big\|^2_{\ell^2}+\big\|\dt \orho_{\boldsymbol{\theta}}(H_i) +\orho_{\boldsymbol{\theta}}(H_i)\, \nu(H_i,I_{\boldsymbol{\theta}}^{(\kappa)}(H_i)) \big\|^2_{\ell^2}\big).
\end{equation}
We note that there is an abuse of notation in the mean squared error of the reconstruction equations.  Technically, $\dt \orho + \orho \nu$ is a differential one-form in $\dt H$.  Thus, when computing the $\ell^2$-norm, we implicitly only consider the components of that one-form.

We supplement the loss function \eqref{lost function} with the initial conditions \eqref{inv initial conditions} and \eqref{orho initial conditions} by considering the invariant initial conditions loss function
\[
\mathcal{L}_{\text{I.C.}}(\boldsymbol{\theta}) = \big\| I_{\boldsymbol{\theta}}^{(k-1)}(H_0) - I_0^{(k-1)}\big\|^2_{\ell^2} + \big\|\orho_{\boldsymbol{\theta}}(H_0) - \orho_0\big\|^2_{\ell^2}.
\]
The final invariant physics-informed loss function is thus given by
\[
\mathcal{L}_{\text{Inv}}(\boldsymbol{\theta}) = \mathcal{L}_{\Delta_{\text{Inv}},\orho}(\boldsymbol{\theta}) + \alpha\, \mathcal{L}_{\text{I.C.}}(\boldsymbol{\theta}), 
\] 
where $\alpha \in\mathbb{R}^+$ is again a hyper-parameter rescaling the importance of the equation and initial condition losses.
  
\section{Examples}\label{sec:Examples}

In this section we implement the invariant physics-informed neural network procedure introduced in Section \ref{sec:InvariantPINNs} for several examples. We also train a standard physics-informed neural network to compare the solutions obtained. For both models we use feed-forward neural networks minimizing the invariant loss function and standard PINN loss function, respectively. For the sake of consistency, all networks used throughout this section have 5 layers, with 40 nodes per layer, and use the hyperbolic tangent as activation functions. For most examples, the loss stabilizes at fewer than 3{,}000 epochs, but for uniformity we trained all models for 3{,}000 epochs. All examples use 200 collocation points.  The numerical errors of the two neural network solutions are obtained by comparing the numerical solutions to the exact solution, if available, or to the numerical solution obtained using {\tt odeint} in \texttt{scipy.integrate}. We also compute the mean square error over the entire interval of integration for all examples together with the standard deviation averaged over $5$ runs. These results are summarized in Table~\ref{tab:errors}. Finally, the point-wise square error plots for each example are provided to show the error varying over the interval of integration. 

\begin{example}\label{ex: Schwarz PINN}
As our first example, we consider the Schwarz equation \eqref{Schwarz}, with $F(t) = 2$.  For the numerical simulations, we used the initial conditions 
\begin{equation}\label{Schwarz initial cond}
u_0 = u_{tt}^0 = 0,\qquad u_t^0 = 1,
\end{equation}
in \eqref{Schwarz initial conditions} with $t_0=0$. According to \eqref{I sol} the invariantization of Schwarz' equation yields the algebraic constraint $I=2$.  Thus, the loss function \eqref{lost function} will only contain the reconstruction equations \eqref{Schwarz reconstruction equations}.  Namely,
\begin{subequations}\label{num schwarz IVP}
\begin{equation}\label{num schwarz reconstruction eq}
\alpha_t +\beta = \beta_t - \alpha = \gamma_t + \delta = \delta_t - \gamma=0,
\end{equation}
where we used the fact that $\sigma = 1$.  Substituting \eqref{Schwarz initial cond} into \eqref{Schwarz reconstruction eq initial cond}, yields the initial conditions 
\begin{equation}
\delta_0 = \alpha_0 = \pm 1,\qquad \beta_0 = \gamma_0 = 0
\end{equation}
\end{subequations}
for the reconstruction equations. In our numerical simulations we worked with the positive sign.  Once the reconstruction equations have been solved, the solution to the Schwarz equation is given by the ratio \eqref{Schwarz sol}.  The solution is integrated on the interval $t\in [0,\pi]$.  Error plots for the solutions obtained via the invariant PINN and the standard PINN implementations are given in Figure \ref{fig:SchPINN}.  These errors are obtained by comparing the numerical solutions to the exact solution $u(t) = \tan(t)$.  Clearly, the invariant implementation is substantially more precise near the vertical asymptote at $t = \pi/2$. The reason for this improvement originates from the fact that the invariant PINN implementation seeks to solve the system of linear ODEs \eqref{num schwarz IVP} to approximate the exact solutions $\beta(t) = \sin(t)$ and $\delta(t) = \cos(t)$, which are bounded functions, while the standard PINN implementation attempts to approximates the unbounded function $u(t) = \tan(t) = \beta(t)/\delta(t)$ using the Schwarz equation \eqref{Schwarz}.


\begin{figure}[!h]
\centering
     \begin{subfigure}[b]{0.49\textwidth}
         \centering
         \includegraphics[width=\textwidth, height=6.25cm]{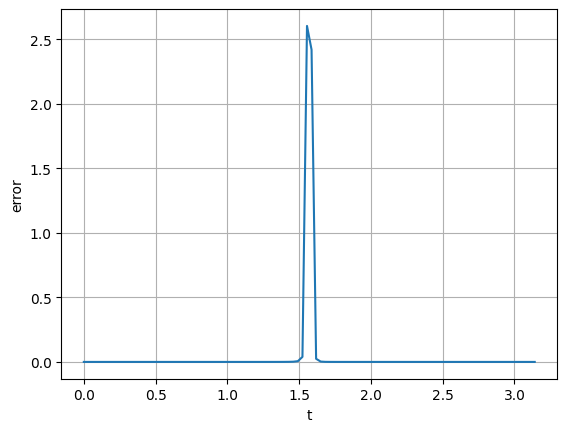}
         \caption{Invariant PINN error.}
     \end{subfigure}
     \hfill
     \begin{subfigure}[b]{0.49\textwidth}
         \centering
         \includegraphics[width=\textwidth, height=6.25cm]{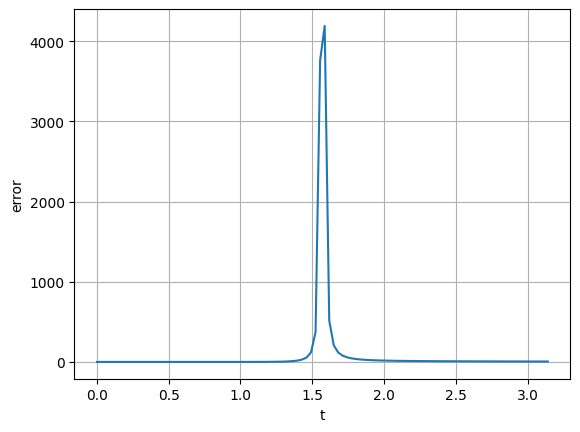}
         \caption{Standard PINN error.}
     \end{subfigure}
     \caption{Time series of the squared error for the Schwarz equation~\eqref{Schwarz}.}\label{fig:SchPINN}
\end{figure}
\end{example}

\begin{example}\label{ex: logistic}
As our second example, we consider the logistic equation
\begin{equation}\label{logistic}
u_t = u(1-u)
\end{equation}
occurring in population growth modeling.  Equation \eqref{logistic} 
admits the one-parameter symmetry group
\[
T=t,\qquad U = \frac{u}{1+\epsilon\, u e^{-t}},\qquad \text{where}\qquad \epsilon \in \mathbb{R}.
\]
Implementing the algorithm outlined in Section \ref{summary}, we choose the cross-section $\mathcal{K} = \{u=1\}$.  This yields the invariantized equation 
\[
I = \iota(u_t) = 0.
\]
The reconstruction equation is
\begin{equation}\label{eq: logistic reconstruction}
\epsilon_t= I = 0,
\end{equation}
subject to the initial condition
\[
\epsilon(t_0) = \bigg(\frac{1-u_0}{u_0}\bigg)e^{t_0},
\]
where $u_0=0.5$ and our interval of integration is $[0,\pi]$. The solution to the logistic equation is then given by
\begin{equation}\label{logistic solution}
u(t) = \frac{1}{1+\epsilon\, e^{-t}}.
\end{equation}
As Figure \ref{fig:LogPINN} illustrates, the error incurred by the invariant PINN model is significantly smaller than the standard PINN implementation, by about a factor of more than 100, when compared to the exact solution \eqref{logistic solution}.

\begin{figure}[!h]
\centering
     \begin{subfigure}[b]{0.49\textwidth}
         \centering
         \includegraphics[width=\textwidth, height=6.25cm]{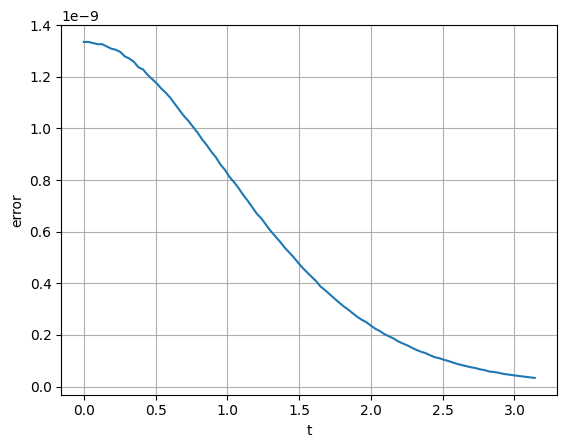}
         \caption{Invariant PINN error.}
     \end{subfigure}
     \hfill
     \begin{subfigure}[b]{0.49\textwidth}
         \centering
         \includegraphics[width=\textwidth, height=6.25cm]{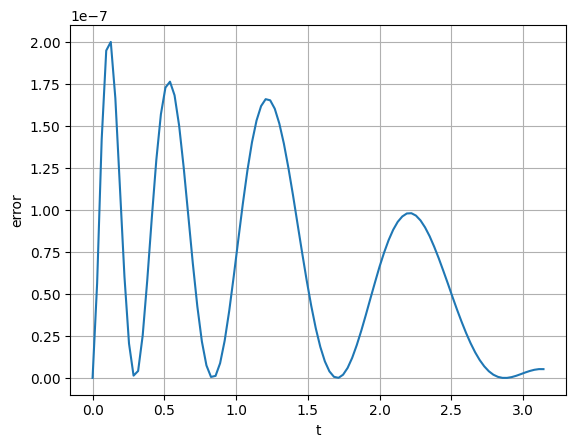}
         \caption{Standard PINN error.}
     \end{subfigure}
     \caption{Time series of the squared error for the logistic equation~\eqref{logistic}. }\label{fig:LogPINN}
\end{figure}
\end{example}

\begin{example}
We now consider the driven harmonic oscillator
\begin{equation}\label{driven oscillator}
u_{tt} + u = \sin(t^a),
\end{equation}
which appears in inductor-capacitor circuits, \cite{Serw03Ay}.  In the following we set $a = 0.99$, which yields bounded solutions close  to resonance occurring when $a = 1$.  The differential equation \eqref{driven oscillator} admits the two-dimensional symmetry group of transformations
\[
T = t,\qquad U = u + \alpha \sin(t) + \beta \cos(t),\qquad \text{where}\qquad \alpha,\beta\in \mathbb{R}.
\]

A cross-section to the prolonged action is given by $\mathcal{K} = \{u=u_t=0\}$.  The invariantization of \eqref{driven oscillator} yields
\[
I = \iota(u_{tt}) = \sin(t^a).
\]
The reconstruction equations are
\begin{equation}\label{driven oscillator reconstruction}
\alpha_t = \sin(t^a) \cos(t),\qquad \beta_t = -\sin(t^a) \sin(t),    
\end{equation}
with initial conditions
\[
\alpha(t_0) = u_0 \sin(t_0) + u_t^0 \cos(t_0),\qquad
\beta(t_0) = u_0\cos(t_0) - u_t^0 \sin(t_0),
\]
where, in our numerical simulations, we set $u_0=u_t^0 = 1$ and integrate over the interval $[0,10]$.  Given a solution to the reconstruction equations \eqref{driven oscillator reconstruction}, the solution to the driven harmonic oscillator \eqref{driven oscillator} is
\[
u(t) = \alpha(t)\, \sin(t) + \beta(t) \cos(t).
\]
Figure \ref{fig:OscillatorPINN} shows the error for the invariant PINN implementation and the standard PINN approach compared to the solution obtained using the Runge--Kutta method {\tt odeint} as provided in {\tt scipy.integrate}.  As in the previous two examples, the invariant version yields substantially better numerical results than the standard PINN method.

\begin{figure}[!h]
\centering
     \begin{subfigure}[b]{0.49\textwidth}
         \centering
         \includegraphics[width=\textwidth, height=6.25cm]{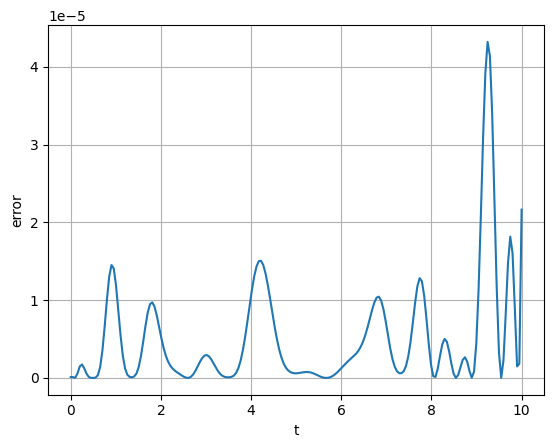}
         \caption{Invariant PINN error}
     \end{subfigure}
     \hfill
     \begin{subfigure}[b]{0.49\textwidth}
         \centering
         \includegraphics[width=\textwidth, height=6.1cm]{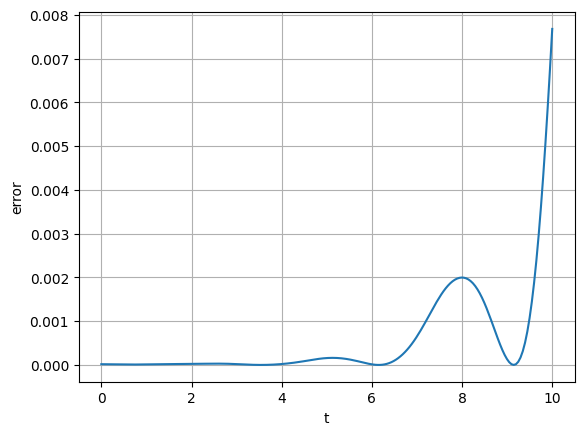}
         \caption{Standard PINN error}
     \end{subfigure}
     \caption{Time series of the squared error for the driven harmonic oscillator~\eqref{driven oscillator}.}\label{fig:OscillatorPINN}
\end{figure}
\end{example}

\begin{example}\label{ex: exponential}
We now consider the second order ordinary differential equation 
\begin{equation}\label{eq:exponential}
u_{tt} = \exp{[-u_t]} 
\end{equation}
with an exponential term.  Equation \eqref{eq:exponential} admits a three-dimensional symmetry group action given by
\[
T = e^{\epsilon}t + a, \qquad  U = e^{\epsilon}u + \epsilon\,  e^{\epsilon} x + b, 
\]
where $a,b,\epsilon \in \mathbb{R}$.  In the following, we only consider the one-dimensional group
\[
T = e^\epsilon t,\qquad U = e^{\epsilon}u + \epsilon\,  e^{\epsilon} x.
\]
We note that in this example the independent variable is not invariant as in the previous examples.  A cross-section to the prolonged action is given by $\mathcal{K} = \{u_t = 0\}$. Introducing the invariants
\[
H = \ln\bigg[\frac{1}{1-\iota(t)}\bigg]=\ln\bigg[\frac{1}{1-t u_t}\bigg],\qquad I=\iota(u)=\exp[-u_t](u-tu_t),
\] 
the invariantization of the differential equation \eqref{eq:exponential} reduces to the first order linear equation
\begin{subequations}\label{eq: exponential foliation}
\begin{equation}
I_H + I = e^{-H}-1.
\end{equation}
The reconstruction equation for the left moving frame is simply
\begin{equation}
\epsilon_H = 1.
\end{equation}
\end{subequations}
In terms of $\epsilon$ and $I$, the parametric solution to the original differential equation \eqref{eq:exponential}  is
\begin{equation}\label{tu exp eq}
t = e^{\epsilon}(1-e^{-H}), \qquad u = e^{\epsilon}(I + \epsilon(1- e^{-H})).
\end{equation}
The solution to \eqref{eq:exponential} is known and is given by
\begin{equation}\label{exact sol}
u(t) = (t+c_1)\ln(t+c_1)-t+c_2,
\end{equation}
where $c_1$, $c_2$ are two integration constants.  For the numerical simulations, we use the initial conditions 
\[
I_0 = \exp[-u_t^0](u_0-t_0u_t^0),\qquad \epsilon_0 = u_t^0,
\]
where $u_0 = u(t_0)$, $u_t^0 = u_t(t_0)$ with $t_0=0$, and $c_1=\exp(-5)$, $c_2=0$ in \eqref{exact sol}. The interval of integration $[H_0,H_f]$ is given by
\begin{equation}\label{H0Hf}
H_0 = \ln\bigg[\frac{1}{1-t_0 u_t^0}\bigg],\qquad H_f = \ln\bigg[\frac{1}{1-t_f u_t^f}\bigg],
\end{equation}
where $u_t^f = u_t(t_f)$ and $t_f=2$.  We choose the interval of integration given by \eqref{H0Hf}, so that when $t$ is given by \eqref{tu exp eq} it lies in the interval $[0,2]$.

Figure \ref{fig:ExpPINN} shows the error obtained for the invariant PINN model when compared to the exact solution \eqref{eq:exponential}, and similarly for the non-invariant PINN model.  As in all previous examples, the invariant version drastically outperforms the standard PINN approach.

\begin{figure}[!ht]
\centering
     \begin{subfigure}[b]{0.49\textwidth}
         \centering
         \includegraphics[width=\textwidth, height=6.25cm]{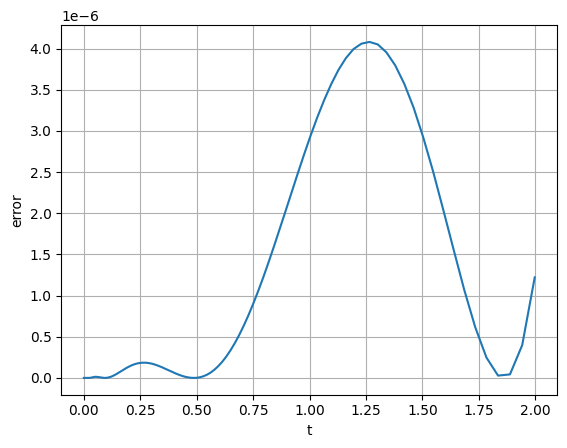}
         \caption{Invariant PINN error.}
     \end{subfigure}
     \hfill
     \begin{subfigure}[b]{0.49\textwidth}
         \centering
         \includegraphics[width=\textwidth, height=6.025cm]{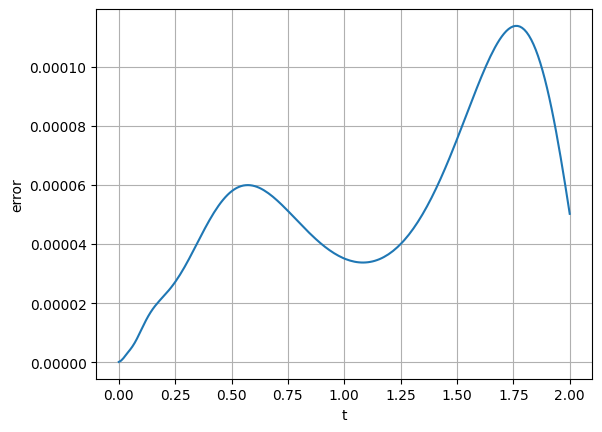}
         \caption{Standard PINN error.}
     \end{subfigure}
     \caption{Time series of the squared error for the exponential equation~\eqref{eq:exponential}.}\label{fig:ExpPINN}
\end{figure}
\end{example}

\begin{example}\label{ex: system}
As our final example, we consider a system of first order ODEs
\begin{equation}\label{system}
    u_t = -u + (t+1)v, \qquad  v_t = u -tv.
\end{equation}
This system admits a two-dimensional symmetry group of transformations given by
\[
T = t,\qquad  U = \alpha u + \beta t,\qquad V = \alpha v + \beta,
\]
where $\alpha > 0$ and $\beta \in \mathbb{R}$.  Working with the cross-section $\mathcal{K} = \{u=1, v=0\}$, the invariantization of \eqref{system} yields
\[
I = \iota(u_x) = -1,\qquad J = \iota(v_x) = 1.
\]
The reconstruction equations are
\begin{equation}\label{eq: system reconstruction}
\alpha_t = \alpha(1+t), \qquad \beta_t = \alpha
\end{equation}
subject to the initial conditions $\alpha_0 = 1$, $\beta_0 = 1$, corresponding to the initial conditions $u_0=v_0=1$, when $t_0=0$.  In our numerical simulations we integrated over the interval $[0,2]$.  The solution to \eqref{system} is then given by
\[
u(t) = \alpha(t) + t\, \beta(t),\qquad v(t) = \beta(t).
\]
As in all previous examples, comparing the numerical solutions to the exact solution
\[
u(t) = \sqrt{\frac{2}{\pi}}\,c\, e^{-(t+1)^2/2} + c\,t\, 
\text{erf}\bigg(\frac{t+1}{\sqrt{2}}\bigg) + kt,\qquad
v(t) = c\, \text{erf}\bigg(\frac{t+1}{\sqrt{2}}\bigg) + k,
\]
with $c=\big(\sqrt{2/\pi}\, \exp(-1/2)\big)^{-1}$ and $k=1-c\, \text{erf}(1/\sqrt{2})$, where
$\text{erf}(t) = 2/\sqrt{\pi}\int_0^t e^{-x^2}\dt x$ is the standard error function, we observe in Figure \ref{fig:SysPINN} that the invariant version of the PINN model considerably outperforms its non-invariant counterpart.

\begin{figure}[!h]
\centering
     \begin{subfigure}[b]{0.49\textwidth}
         \centering
         \includegraphics[width=\textwidth, height=6.25cm]{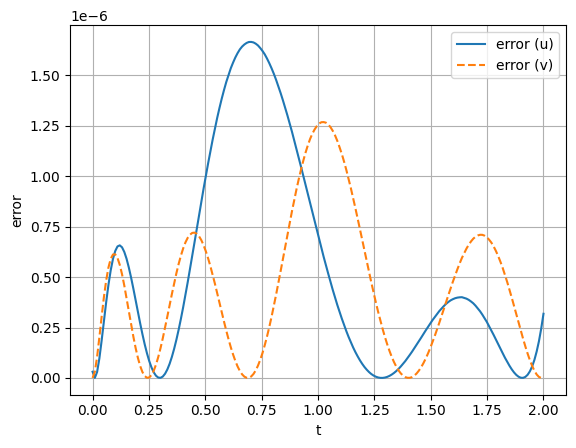}
         \caption{Invariant PINN error.}
     \end{subfigure}
     \hfill
     \begin{subfigure}[b]{0.49\textwidth}
         \centering
         \includegraphics[width=\textwidth, height=6.025cm]{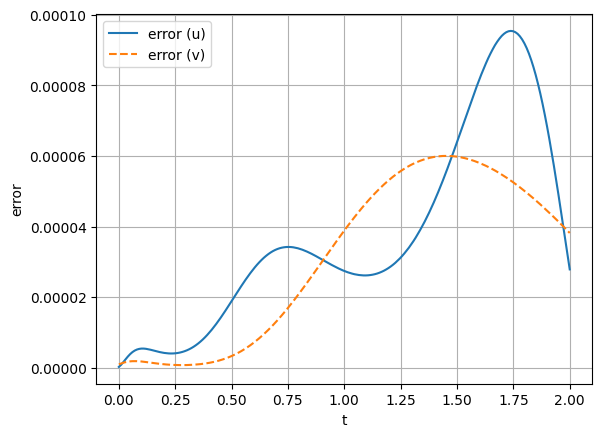}
         \caption{Standard PINN error.}
     \end{subfigure}
     \caption{Time series of the squared error for the system of equations \eqref{system}.}
      \label{fig:SysPINN}
\end{figure}
\end{example}

\section{Summary and conclusions}\label{sec:Conclusion}

In this paper we have introduced the notion of invariant physics-informed neural networks. These combine physics-informed neural networks with symmetry methods for differential equations to simplify the form of the differential equations that have to be solved. In turn, this simplifies the loss function that has to be minimized. For example, in the case of the Schwarz equation considered in Examples \ref{ex: Schwarz PINN}, the third order nonlinear differential equation \eqref{Schwarz} with $F(t)=2$, is replaced by the system of first order linear reconstruction equations \eqref{num schwarz reconstruction eq}.  Similarly, the  nonlinear logistic equation \eqref{logistic} in Example \ref{ex: logistic} is replaced by the linear reconstruction equation \eqref{eq: logistic reconstruction}.  The same phenomenon occurs in Example \ref{ex: exponential}, where the nonlinear equation \eqref{eq:exponential} is substituted by the system of linear equations \eqref{eq: exponential foliation}.  Finally, in Example \ref{ex: system} the coupled system of linear differential equations \eqref{system} is replaced by the triangular system of differential equations \eqref{eq: system reconstruction}. Our numerical tests show that the solutions obtained with the (simplified) invariant models outperformed their non-invariant counterparts.  Table~\ref{tab:errors} summarizes the examples considered in the paper and shows that the invariant PINNs drastically improve over vanilla PINNs for all examples considered.  Quoting \cite{perlis82} ``Symmetry is a complexity-reducing concept $\ldots$" which in the context of the current paper is, we conjecture, at the source of the numerical improvements.

\begin{table}[!ht]
\renewcommand{\arraystretch}{1.2}
\begin{center}
\begin{tabularx}{0.8\textwidth} { 
  | >{\centering\arraybackslash}X 
  || >{\centering\arraybackslash}X 
  | >{\centering\arraybackslash}X 
  | >{\centering\arraybackslash}X | }
 \hline
 \textbf{Example}  & \textbf{Invariant PINN} & \textbf{Vanilla PINN} \\
 \hline\hline
Schwarz (\ref{Schwarz})  & $5.1 \cdot 10^{-2} \pm 4.1 \cdot 10^{-2}$  & $96.89 \pm 0.81$ \\
  \hline
 Logistic (\ref{logistic})   & $2.6 \cdot 10^{-10} \pm 1.8 \cdot 10^{-10}$  & $3\cdot 10^{-2} \pm 4.6\cdot 10^{-5}$ \\
\hline  
Harmonic (\ref{driven oscillator}) & $5.8 \cdot 10^{-6} \pm 4.8 \cdot 10^{-6}$   & $1.4\cdot 10^{-5} \pm 2.3\cdot 10^{-5}$ \\
  \hline
  Exponential (\ref{eq:exponential})  & $3.2 \cdot 10^{-7} \pm 2.2 \cdot 10^{-7}$  & $2.6\cdot 10^{-5} \pm 3.6\cdot 10^{-5}$ \\
  \hline
System \eqref{system} & $9.4\cdot 10^{-7} \pm 2.0\cdot 10^{-7} $ & $6.6\cdot 10^{-6} \pm 7.9\cdot 10^{-6}$ \\
  \hline
\end{tabularx}
\end{center}
\caption{Mean square error with standard deviation averaged over five runs for all examples considered in Section~\ref{sec:Examples}.}\label{tab:errors}
\end{table}

The proposed method is fully algorithmic and as such can be applied to any system of differential equations that is strongly invariant under the prolonged action of a group of Lie point symmetries.  It is worth noting that the work proposed here parallels some of the work on invariant discretization schemes which, for ordinary differential equations, also routinely outperform their non-invariant counterparts. We have observed this to also be the case for physics-informed neural networks.

Lastly, while we have restricted ourselves to the case of ordinary differential equations, our method extends to partial differential equations as well.  Though, when considering partial differential equations, it is not sufficient to project the differential equation onto the space of differential invariants as done in this paper.  As explained in \cite{thom15Ay}, integrability conditions among the differential invariants must also be added to the invariantized differential equation.  In the multivariate case, the reconstruction equations \eqref{reconstruction eq} will then form a system of first order partial derivatives for the left moving frame.  Apart from these modifications, invariant physics-informed neural networks can also be constructed for partial differential equations, which will be investigated elsewhere.

\section*{Acknowledgement}
This research was undertaken, in part, thanks to funding from the Canada Research Chairs program and the NSERC Discovery Grant program. The authors also acknowledge support from the Atlantic Association for Research in the Mathematical Sciences (AARMS) Collaborative Research Group on \textit{Mathematical Foundations for Scientific Machine Learning}.

{\footnotesize\setlength{\itemsep}{0ex}
\bibliography{bihlo}}

\end{document}